\documentstyle[preprint,prb,aps]{revtex}
\tightenlines
\begin{document}
\draft
\title{The Functional Form of Angular Forces around Transition Metal Ions in Biomolecules}
\author{A. E. Carlsson and S. Zapata}
\address{Department of Physics\\
Washington University\\
St. Louis, Missouri 63130-4899}
\date{Submitted to Biophysical Journal, 28 February 2000}
\maketitle
\begin{abstract}
A method for generating angular forces around $\sigma$-bonded transition 
metal ions is generalized to treat $\pi$-bonded configurations. 
The theoretical approach is based on an analysis of  a ligand-field
Hamiltonian based on the moments of the electron state distribution. 
The functional forms that are obtained involve a modification of
the usual expression of the binding energy as a sum of ligand-ligand 
interactions, which however requires very little increased in CPU time. 
The angular interactions have simple forms involving sin and cos 
functions, whose relative weights depend on whether the ligands are 
$\sigma$- or $\pi$-bonded. They describe the ligand-field stabilization
energy to an accuracy of about 10\%.  The resulting force field 
is used to model the structure of small clusters, including fragments 
of the copper blue protein structure. Large deviations from the typical
square copper coordination are found when $\pi$-bonded ligands are
present.
\end{abstract}
\pacs{Running title:  Angular Forces around IonsForce fields, Metal complexes}

\newpage

\section{Introduction}

Atomic-level simulations of proteins interacting with transition-metal 
ions have the promise of elucidating a wide variety of biophysical 
%%phenomena\cite{Kaim94}. 
phenomena~(Kaim, 1994). 
%For example, the function of both redox and non-redox enzymes often 
%is centered on one or more metal ions. In these cases, simulations
%can elucidate the underlying structure and how it response to substrate
%molecules or to changes in the charge of the metal ions. Metal ions
%often served as structural ``glue" in stabilizing the tertiary structure
%of folded proteins (such as zinc 
%%fingers\cite{}), 
%fingers~(?..., 19..?)), and the secondary
%structure of shorter peptides. Simulations can aid in the prediction of
%the extent and specificity of such ``glue" effects. 
Such simulations can help establish both the native structure and
reaction paths of metalloproteins, and explain the roles of metal
ions in protein folding and stability. Similar applications can
be seen for simulations of metal ions interacting with DNA and RNA. 
However, the utility of such simulations depends critically on the 
availability of accurate but computationally tractable force fields 
for metals interacting with proteins. There are no quantitatively 
accurate force fields for these purposes, and it is probably not 
possible to construct force fields of a simple enough form to be 
useful that have quantitative accuracy. 
However, one can hope to generate force fields that include the basic
physical effects at a level of accuracy sufficient to ascertain
chemical trends as metals or ligands vary. The situation in this regard 
is best for ``simple" metal ions that have no partly occupied shells. 
Here an ionic picture based on electrostatics, supplemented by empirical 
repulsive forces, may hope to treat the most important physical effects. 
Even here though, care should be taken, since the bond is not completely 
ionic but has some partial covalent character. For transition metals, 
the situation is much more difficult. The ligand-field splitting
of the $d$-shell leads to important electronic contributions which 
cannot be ignored. These are manifested, for example, in the typical 
square or tetragonal coordination of Cu$^{2+}$ complexes. 
Such effects cannot be described by radial interactions, 
but rather require the introduction of angular forces describing 
the energetics of the transition-metal $d$-shell. 
Most existing methods for including angular forces in simulations 
of metal ions have used assumed functional 
%%forms\cite{Comba95,Allured91,Timofeeva95,Wiesemann94,Sayle95,Sayle97}
forms~(Comba et al., 1995, Allured et al., 1991, Timofeeva et al., 1995,
Wiesemann et al., 1994, Sayle et al., 1995, Sayle et al., 1997)
based, for example on the observed structures of small complexes. 
A recently developed 
%method\cite{Landis98} 
method~(Landis et al., 1998) 
treats the environment-dependent energetics of transition metals 
via a valence-bond approach. This method appears promising for 
covalently bonded metals. However, the bonding state of transition 
metals in proteins is may be rather different from the $sd^n$ 
hybridized configuration assumed in that work. In the present analysis, 
we seek to develop a method suitable for more ionic configurations. 

We have recently 
%shown\cite{Carlsson98} 
shown~(Carlsson, 1998) how the ligand-field splitting effects of 
$\sigma$-bonded transition metal ions can be described by a force 
field of a fairly simple form. One starts with an explicitly 
quantum-mechanical form for the electronic energy, and then then 
solves the quantum-mechanical problem with systematic approximation 
methods to extract the real-space description of the electronic
bonding energy.  A second-order treatment of the hybridization 
terms between the ligand and metal orbitals is used to generate 
a ligand-field Hamiltonian for the $d$-electrons. The electronic 
bonding energy of this Hamiltonian is analyzed in terms its moments. 
The moment analysis of the energy gives a ``semiclassical" energy 
function with a simple trigonometric angular form. 
It is not precisely an additive function of ligand-ligand interactions, 
but is nearly as simple computationally. This energy function gives 
quite accurate energy results for $\sigma$-bonded ligands, 
with less than 10\% error in the ligand-field stabilization energy 
(defined below). 

This approach is extended in the present paper, in two ways. 
First, we develop energy functions which treat $\pi$-bonded
ligands as well. This results in new and distinct angular forms. 
Second, we present results for the lowest-energy structures 
of small clusters, which are a guide to understanding the bonding 
preferences of transition metals in proteins.

The organization of the remainder of the paper is as follows. 
Section~II develops the formalism underlying the angular
forces. Section~III gives tests of the functional form of
the angular forces by comparing with results from diagonalization
of simple cluster Hamiltonians. Section~IV gives the small-cluster
results for ideal examples, and for an analogue of the copper
environment in blue-copper proteins.

\section{Model}

We model the electronic structure of the environment of a single
transition metal ion in terms of orbitals $|L,\mu\rangle$ 
localized on the ligands and $|M,\nu\rangle$ localized on the  
single metal ion. The orbitals taken to be orthogonal for simplicity 
of calculation. Here ``L" denotes a particular ligand atom and
the $|L,\mu\rangle$ are distinct orbitals on that atom. 
The Hamiltonian takes the following form:
\begin{eqnarray}
  \hat H =
&& \sum_{L,\mu} \varepsilon_{L,\mu} |L,\mu \rangle\langle L,\mu | 
    +\sum_\nu \varepsilon_{M,\nu} 
     |M,\nu \rangle\langle M,\nu |\nonumber \\
%+ && \sum_{L,\mu ,\nu } [h_{LM}^{\mu \nu} |L,\mu ><M,\nu | +
   && + \sum_{L,\mu ,\nu} \Bigl[ h_{LM}^{\mu\nu} 
      |L,\mu \rangle\langle M,\nu | + h_{ML}^{\nu \mu} 
      |M,\nu \rangle\langle L,\mu | \Bigr]\quad .
\label{eqnarray}
\end{eqnarray}
Here the $\varepsilon$ terms are on-site energies for the orbitals, 
and the $h$ terms are hybridization energies between the ligand orbitals
and the metal orbitals. We ignore electron-interaction terms as
well as explicit electrostatic effects. The energy associated with
this Hamiltonian can then be obtained by diagonalizing its matrix
as given in the $|L,\mu \rangle -|M,\nu\rangle$ basis,
and taking the sum of the energies of the occupied eigenvectors.  

To obtain a real-space description of the energetics of this Hamiltonian, 
we make the simplifying approximation that the dimensionless ratios
$|h_{LM}^{\mu \nu}|/(\varepsilon_{M,\nu}-\varepsilon_{L,\mu})$ are small
and can be used as expansion parameters. This will be the case if the
bonding is primarily ionic as opposed to covalent. In the case of
transition metal ions, this allows a great simplification by the 
application of ligand-field theory, which gives an effective 
Hamiltonian for the $d$-shell: 
\begin{equation}
     \hat H_{d} 
   = \sum_{\mu ,\nu}h_d^{\mu\nu} |M,\mu \rangle\langle M,\nu |\quad ,
\label{hd}
\end{equation}
where
\begin{equation}
   h_d^{\mu\nu} 
 = \sum_{L,\eta}(\varepsilon_d - \varepsilon_{L,\eta })^{-1}
   h_{ML}^{\mu \eta} h_{LM}^{\eta \nu}\quad .
\label{hlft}
\end{equation} 
Note that we have assumed all of the $d$-orbitals on the transition metal
ion to have the same energies before ligand-field effects are ``turned on". 

In a fully quantum-mechanical ligand-field-theory calculation, 
one would numerically diagonalize the matrix $\hat H_d$. 
In order to obtain a force field with a nearly classical form, 
we instead work with the moments of this matrix,
defined in terms the traces of its powers, as follows. 
The first moment is defined as the average energy of the $d$-complex, or
\begin{eqnarray}
   \bar \varepsilon 
&=& (1/5) {\rm Tr}\hat H_d 
    =(1/5)\sum_{L,\eta}(\varepsilon_d -\varepsilon_{L,\eta})^{-1}
     \sum_\mu h_{ML}^{\mu \eta} h_{LM}^{\eta \mu} \nonumber  \\
&=& (1/5)\sum_{L,\eta}(\varepsilon_d -\varepsilon_{L,\eta})^{-1} 
    g_{\eta\eta}
\label{mu1}
\end{eqnarray}
where
\begin{equation}
   g_{\eta ,\eta '}=\sum_\mu h_{L'M}^{\eta '\mu}h_{ML}^{\mu\eta}\quad . 
\label{g}
\end{equation}
%In order to get a better feeling for these terms, we note 
%that $g_{\eta\eta}=\langle L,\eta |\hat P_d | L,\eta\rangle$, where
%$\hat P_d=\sum_\mu |M,\mu \rangle\langle M,\mu |$ is the projection onto the
%$d$-subspace of the transition metal ion. Since $\hat P_d$ is a
%rotationally invariant operator, it follows that $g_{\eta \eta}$
%is a radial function of the distance between the ligand L and the metal ion. 
Thus $\bar\varepsilon$ is given as a sum of independent contributions 
from the ligands. Because the $d$-shell by itself has spherical symmetry, 
the contribution from each ligand is a radial function (no angular 
dependence) of the metal-ligand distance. 
Our major interest is in the angular terms resulting from the
ligand-field splitting, so we do not consider the $\bar\varepsilon$
term further. 

The width $W$ of the $d$-complex corresponding to the ligand-field 
splitting can in the simplest picture be described in terms of 
the second moment or variance $\delta\varepsilon^2$ of the 
eigenvalues of $\hat H_d$. We expect that 
$W\propto\sqrt{\delta\varepsilon^2}$. We can obtain a simple
real-space form for $\delta\varepsilon^2$ as follows: First we 
note that 
$\delta\varepsilon^2=(1/5)\sum_n(\varepsilon_n-\bar\varepsilon)^2
=(1/5){\rm Tr}(\hat H_d-\bar\varepsilon\hat I )^2$
where the $\varepsilon_n$ are the eigenvalues of $\hat H_d$.
From Eqs.~(\ref{hd}), (\ref{hlft}), and (\ref{g}), we see that 
\begin{eqnarray}
     {\rm Tr}\hat H_d^2 
  &=& \sum_{\mu ,\nu}h_d^{\mu\nu} h_d^{\nu\mu} \nonumber \\ 
  &=& \sum_{L,\eta ,L',\eta '}(\varepsilon_d-\varepsilon_{L,\eta})^{-1}
      (\varepsilon_d-\varepsilon_{L',\eta '})^{-1} 
      \sum_\mu h_{L'M}^{\eta ' \mu}h_{ML}^{\mu \eta}
      \sum_\nu h_{LM}^{\eta\nu}h_{ML'}^{\nu\eta '} \nonumber \\
  &=& \sum_{L,\eta ,L',\eta '}(\varepsilon_d-\varepsilon_{L,\eta})^{-1}
      (\varepsilon_d-\varepsilon_{L',\eta '})^{-1} g_{\eta \eta '}^2\quad .
\end{eqnarray}
Then the variance is
\begin{eqnarray}
      \delta\varepsilon^2 
  &=& (1/5){\rm Tr}\Bigl( \hat H_d-\bar\varepsilon \hat I\Bigr)^2 
      =(1/5)\Bigl( {\rm Tr}\hat H_d^2 -5\bar\varepsilon^2\Bigr)\nonumber \\
  &=& (1/5)\sum_{L,\eta ,L', \eta '} 
      (\varepsilon_d-\varepsilon_{L,\eta})^{-1}
      (\varepsilon_d-\varepsilon_{L',\eta '})^{-1} 
      \Bigl[ g_{\eta \eta '}^2-(1/5)g_{\eta\eta} 
             g_{\eta ',\eta '}\Bigr]\quad .
\label{variance}
\end{eqnarray}

We will focus on the ligand-field stabilization energy, which is
defined as the sum of the energies of the occupied orbitals
relative to $\bar\varepsilon$, or 
$E_{\rm LFSE}=\sum_{n}' (\varepsilon_n -\bar\varepsilon )$,
where the sum is over only the occupied eigenfunctions.
We assume that $E_{\rm LFSE}$ is proportional to the ligand-field 
splitting $W$, and also identify $W$ with  the standard deviation 
$\sqrt{\delta \varepsilon^2}$.
Thus our expression for the ligand-field stabilization energy
has the form
\begin{equation}
    E_{\rm LFSE} =  -\biggl\{ \alpha /5 
    \sum_{L,\eta ,L', \eta '} 
    (\varepsilon_d-\varepsilon_{L,\eta})^{-1}
    (\varepsilon_d-\varepsilon_{L',\eta '})^{-1} 
    \Bigl[ g_{\eta\eta '}^2-(1/5) g_{\eta\eta} 
           g_{\eta ',\eta '}\Bigr]\biggr\}^{1/2}\quad ,
\label{elfse}
\end{equation}
where $\alpha$ is a dimensionless constant which will later be
used as a fitting parameter.

The accuracy of this form will be demonstrated in the next section
via specific numerical experiments. At present, we will show how
this form for $E_{\rm LFSE}$ results in an expression for the energy
in terms of simple angular interactions between he ligands. 
We thus need to develop analytic forms for the $g_{\eta\eta '}$.
First we note that we can write
\begin{eqnarray}
      g_{\eta \eta '} 
  &=& \langle L,\eta | \hat H P_d^2 \hat H |L', \eta '\rangle \nonumber \\
  &=& \langle \Psi | \Psi '\rangle \quad ,
\end{eqnarray}
where $\hat P_d = \sum_\mu |M,\mu\rangle\langle M,\mu |$ 
is the projection onto the $d$-subspace of the transition metal ion
Here we define $|\Psi \rangle =\hat P_d H |L,\eta\rangle$  
and we have have used the relation $\hat P_d^2 = \hat P_d$, which
follows from $\hat P_d$'s being a projection operator. 

Consider first the case where both the $|L,\eta\rangle$ and 
$|L',\eta '\rangle$ orbitals have $\sigma$-character with respect to  
the metal ion. Then they couple only to the $d$-orbitals 
$|M,\sigma\rangle$ and $|M,\sigma '\rangle$ that have 
$\sigma$-character with respect to the bond axes, so that 
$|\Psi\rangle = h_\sigma |M, \sigma\rangle$ and
$|\Psi '\rangle = h_{\sigma }'|M, \sigma ' \rangle$, where
$h_\sigma$ and $h_{\sigma}'$ are the appropriate coupling strengths.
Thus 
$g_{\eta\eta '}=h_\sigma h_{\sigma}'\langle M,\sigma | M,\sigma '\rangle$,
and the angular dependence is contained in the last inner product. 
But this is simply a matrix element of a rotation about $M$ which 
carries $L$ into $L'$. We note that $|M, \sigma\rangle$ is equivalent 
to $|M,m=0\rangle$, where $m=0$ denotes the angular dependence of
the spherical harmonic $Y_{2m}$.
Choosing our coordinate system so that $L$ is along the
$z$-axis and $L'$ is in the $z$-$x$ plane, we find that
\begin{eqnarray}
     \langle M,\sigma | M,\sigma '\rangle 
 &=& D_{00}^{(2)}(0,\theta , 0) \nonumber \\
 &=& (3 \cos^2 \theta -1)/2
\end{eqnarray}
where the $D$-term is a matrix element of the $l=2$ representation
of the rotation group, and the second equality follows from the
explicit formulas of the $D$-terms given in 
%%Ref.\onlinecite{Wigner59}.
Ref.~(Wigner, 1959).
In summary, for $\sigma$-bonded ligands, 
\begin{equation}
   g_{\eta\eta `}=h_\sigma h_{\sigma}' (3\cos^2\theta -1)/2\quad .
\label{gsigma}
\end{equation}

For the case of a $\pi$-bonded orbital $L$ and a $\sigma$-bonded 
orbital $L'$, we consider only the case in which the axis of the orbital
lies along the circle connecting $L$ and $L'$; if it is perpendicular
to this circle, $L$ and $L'$ have different inversion symmetries so
their coupling vanishes. We write 
$| \Psi\rangle =h_{\pi} | M,\pi\rangle$ where 
$|M,\pi\rangle =(1/\sqrt{2})(-|M,m=1\rangle +|M, m=-1\rangle )$ and
$m$ the usual aximuthal angular momentum index for the spherical harmonics. 
(The ``$-$" sign comes from the definition of the spherical harmonics).  
Then, again choosing a coordinate system in which $L$ is along the 
$z$-axis and $L'$ is in the $z$-$x$ plane, and following reasoning 
parallel to the $\sigma$-bonded case, we see that
\begin{eqnarray}
      \langle M,\pi | M,\sigma '\rangle  
  &=& (1/\sqrt{2}) \Bigl[ -D_{10}^{(2)}(0,\theta ,0) 
      + D_{-10}^{(2)}(0,\theta ,0)\Bigr]  \nonumber \\
  &=& \sqrt{3} \sin{\theta}\cos{\theta}\quad .
\end{eqnarray}
and
\begin{equation}
   g_{\eta\eta `}=\sqrt{3}h_{\pi} h_{\sigma}'\sin{\theta}\cos{\theta}\quad .
\label{gpisigma}
\end{equation}

Finally we turn to the case of two $\pi$-bonded orbitals. 
To obtain the subsequent results, it is sufficient to consider the
case~1) in which the orbitals are parallel to the arc connecting
$L$ and $L'$, and the case~2) in which they are perpendicular to it. 
By reasoning similar to that above, one sees that in case~1)
\begin{eqnarray}
      \langle M, \pi | M, \pi '\rangle 
  &=& (1/2)\Bigl[ D_{11}^{(2)}(0,\theta ,0) +D_{-1-1}^{(2)}(0,\theta ,0) 
      - D_{1-1}^{(2)}(0,\theta ,0) 
      - D_{-11}^{(2)} (0,\theta ,0)\Bigr]  \nonumber \\
  &=& \cos{2\theta}
\end{eqnarray}
and
\begin{equation}
     g_{\eta\eta `} = h_{\pi} h_{\pi}' \cos{2\theta}\quad .
\label{gpipia}
\end{equation}
In case 2),
\begin{eqnarray}
      \langle M,\pi | M,\pi '\rangle 
  &=& (1/2) \Bigl[ D_{11}^{(2)}(0,\theta ,0) 
      + D_{-1-1}^{(2)}(0,\theta ,0) 
      + D_{1-1}^{(2)}(0,\theta ,0)l
      + D_{-11}^{(2)}(0,\theta ,0)\Bigr]  \nonumber \\
  &=& \cos{\theta}
\end{eqnarray}
and
\begin{equation}
   g_{\eta\eta `} = h_{\pi} h_{\pi}' \cos{\theta}\quad .
\label{gpipib}
\end{equation}

Thus, combining Eqs.~(\ref{elfse}), (\ref{gsigma}), (\ref{gpisigma}),
(\ref{gpipia}), and (\ref{gpipib}), we find the following explicit
form for $E_{\rm LFSE}$ as a sum of ligand-ligand interactions:
\begin{equation}
    E_{\rm LFSE} 
  = -\sqrt{\sum_{L,\eta ,L',\eta '}} U(L,\eta ;L',\eta ')
\label{efinal}
\end{equation}
where the ligand-ligand interaction $U$ is given as follows:
\bigskip

\noindent{For $\sigma$-$\sigma$ interactions,}
\begin{equation}
~~~~~~~~U(L,\eta ;L',\eta ')
        = (\alpha /5) e_{\sigma}(r) e_{\sigma}(r') u_{\sigma\sigma}(\theta )
%[(h_{\sigma}^2
%/(\varepsilon_d - \varepsilon_{L,\eta })]
%[{h_{\sigma}'}^2 /(\varepsilon_d - \varepsilon_{L,\eta '})]
%(9\cos^4{\theta}-6\cos^2{\theta}+1/5)/4,
\label{usigma}
\end{equation}
\noindent{for $\pi$-$\sigma$ interactions,}
\begin{equation}
~~~~~~~~U(L,\eta ;L',\eta ')
        = (\alpha  /5)e_{\pi}(r) e_{\sigma}(r') u_{\pi\sigma}(\theta )
%[(h_{\pi}^2
%/(\varepsilon_d - \varepsilon_{L,\eta })]
%[{h_{\sigma}'}^2 /(\varepsilon_d - \varepsilon_{L,\eta '})]
%(-3\cos^4{\theta}+3\cos^2{\theta}-2/5),
\label{upisig}
\end{equation}
\noindent{and for $\pi$-$\pi$ interactions,}
\begin{equation}
~~~~~~~~U(L,\eta ;L',\eta ')
        = (\alpha /5) e_{\pi}(r) e_{\pi}(r') u_{\pi\pi}(\theta )
%[(h_{\pi}^2
%/(\varepsilon_d - \varepsilon_{L,\eta })]
%[{h_{\pi}'}^2 /(\varepsilon_d - \varepsilon_{L,\eta '})]
%(4\cos^4{\theta}-3\cos^2{\theta}+1/5),
\label{upipi}
\end{equation}
where
\begin{eqnarray}
      e_{\sigma}(r) 
   = && {h_{\sigma}}^2 /(\varepsilon_d - \varepsilon_{L,\eta }), \\
      e_{\pi}(r) 
   = && {h_{\pi}}^2 /(\varepsilon_d - \varepsilon_{L,\eta '}), \\
        u_{\sigma\sigma}(\theta) 
   = && (9\cos^4{\theta}-6\cos^2{\theta}+1/5)/4, \\
        u_{\pi\sigma}(\theta ) 
   = && (-3\cos^4{\theta}+3\cos^2{\theta}-2/5), \\
        u_{\pi\pi}(\theta ) 
   = && (4\cos^4{\theta}-3\cos^2{\theta}+1/5)\quad .
\label{eandu}
\end{eqnarray}
In each case, $U$ is given as a product of radial terms involving
the two ligands, and a simple angular function. (Note that although
denoted an interaction here, $U$ does not have units of energy  because
of the square root in Eq.~(\ref{efinal})). In the calculation of the
terms involving $\pi$-$\sigma$ interactions, each term involves 
a sum over two $\pi$-orbitals, parallel and perpendicular to the
arc connecting the two ligands. In the latter case, $g_{\eta\eta '}$
in Eq.~(\ref{elfse}) vanishes, but $g_{\eta\eta}$  and 
$g_{\eta ' \eta '}$ do not. In the calculation of the terms involving 
$\pi$-$\pi$ interactions, one has a similar scenario except that one 
sums over two pairs of $\pi$-orbitals. 

These forms are plotted out in Figure~1a. Note that
the $\sigma$-$\sigma$ interactions are fairly similar in form 
to the $\pi$-$\pi$ interactions, both having pronounced minima
at $180^{\circ}$ (as well as the physically irrelevant one at
$0^{\circ}$), and a shallower minimum at $90^{\circ}$. The
$\sigma$-$\pi$ interaction is complementary to these,
having minima at $45^{\circ}$ and $135^{\circ}$. We shall
see later that these differences lead to large differences
in ground-state structures of small clusters. 
For comparison, we show in Figure 1b an empirical angular interaction 
%curve\cite{Comba95} 
curve~(Comba et al., 1995) 
assumed in some previous calculations of small transition-metal structures. 
The angular dependence is based on the observed square structure of small 
complexes of the transition metals of interest, and is proportional to
$\sin^2{2\theta}$. Since the metals which have square coordination
generally have predominantly $\sigma$-bonds to their neighbors, 
the most relevant comparison is to the $\sigma$-$\sigma$ curve in Figure~1a. 
We see that the behavior is quite different. The empirical curve has 
equivalent minima at $180^{\circ}$ and $90^{\circ}$, while in the 
theoretical curve the $180^{\circ}$ minimum is much deeper. 

\newpage
\section{Tests of Functional Form}

In order to evaluate the accuracy of the semiclassical form of
Eq.~(\ref{elfse}) for the ligand-field stabilization energy, we
have performed explicit tests for small clusters. These clusters
consist of a central transition-metal ion with four neighbors
placed at random orientations at random distances relative to
the central ion. 
We consider only the minority-spin orbitals, as these determine
$E_{\rm LFSE}$ for high-spin late transition metals.  Four
electrons are placed in these states, corresponding to
Cu$^{2+}$; other band filling values gives similar results.
The random distances are taken into account by
varying the couplings $h_{\sigma}$ and $h_{\pi}$ uniformly over
a finite interval ranging from zero to $h_{\rm max}^\sigma$ or 
$h_{\rm max}^\pi$. We take $h_{\rm max}^\pi = 0.5h_{\rm max}^\sigma$.
The energies are obtained by explicit diagonalization of a 
tight-binding Hamiltonian for this cluster. All of the ligand 
orbitals are taken to have the same value of 
$\varepsilon_{L,\eta}$.  The only dimensionless
variable that enters the results is then 
$\gamma = |h_{\rm max}/(\varepsilon_d -\varepsilon_{L,\eta})|$.
For small values of $\gamma$, the bonding is primarily ionic,
and for larger values it acquires more covalent character. 
We use $\gamma = 0.1$ for our results, but even when $\gamma$ is
significantly larger we find that the accuracy of the
semiclassical form is essentially the same.

We have fitted the energies of these clusters to the following 
semiclassical form for the energy:
\begin{equation}
    E_{\rm LFSE}^2 
 = {\sum_{ L,\eta ,L',\eta '}} U(L,\eta ;L',\eta ') +\beta \quad ,
\label{efit}
\end{equation}
involving two parameters $\alpha$ and $\beta$ (where $U$ contains
$\alpha$). We use a database
of 10,000 clusters to determine the parameters, and then test them on
a set of 1000 clusters not included in the ``training" set. Typical
results are shown in Fig.~2, which shows the semiclassical energies
vs. exact energies for 1000 clusters. Figure~2a corresponds to
three $\sigma$-bonded ligands and one $\pi$-bonded one, while Fig.~2b
corresponds to four $\pi$-bonded ligands. The rms errors in 
$E_{\rm LFSE}$ for these two cases are 9\% and 14\%, respectively.
For clusters with two and three $\pi$-bonded ligands, the rms
errors are 10\% and 11\%. This is to be compared with rms errors 
of 25\%--30\% that are obtained with empirical force 
%%fields\cite{Carlsson98}.
fields~(Carlsson, 1998).

\newpage
\section{Small Cluster Minimum-Energy Structures}

In this section, we describe some of the implications
of the angular forms developed above for the structure 
of small model clusters consisting of a metal atom and
four ligands.  Such cluster calculations cannot treat the protein
environment  accurately; this would await parametrization and 
incorporation of the force field into protein codes, which
is in progress in our group. However, from the small-cluster
calculations it is possible to see the structural preferences
of the angular forces by themselves. These are an important
factor in the final structure adopted by a protein.  

To keep the calculations as simple as possible, we place the
ligands at frozen bond lengths from the central metal ion. 
Here ``frozen means" that they are fixed at a given set of values,
but these values are not necessarily the same for all of the ligands. 
The energy terms include the ligand-field stabilization energy as 
described by our angular terms, as well as a repulsive radial interaction
between ligands forbidding close approaches. The latter term is needed 
because otherwise spurious structures involving
$45^{\circ}$ bond angles can appear when $\pi$-bonding ligands
are present. For simplicity, we assume that the values of
$h_{\sigma}$ and $h_{\pi}$ at a given distance are the same. 
To evaluate the magnitudes of these couplings, we assume
a value of 100~kJ/mole~$=23.9$~kcal/mole~$= 1.04$~eV for
the ligand-field stabilization energy of a cluster with
four $\sigma$-ligands in square coordination; this number
would correspond to the more strongly stabilized 
%complexes\cite{Cotton72}.
complexes~(Cotton, 1972).
For clusters with unequal bond lengths, we assume 
an exponential decay, so that
$h_{\sigma}(r)=h_{\sigma}(r_0)\exp{[-\kappa (r-r_0)]}$,
where $r_0$ is the reference distance and $\kappa$ is a decay parameter. 
Since the most prominent case that we consider of a distant ligand is 
sulfur, we identify $\kappa$ with the spatial decay rate coming from 
the measured first ionization energy $E_1$ of sulfur, 
10.4~eV (=~240~kcal/mole), using the formula 
$\hbar^2 \kappa^2/2m = E_1$. This yields
$\kappa = 1.65${\AA}$^{-1}$.  The ligand-ligand terms 
contain an exponential term taken as the repulsive part
of the van der Waals interactions as given in the ``MM2" 
%force field\cite{Allinger87}. 
force field~(Sprague et al., 1987). 
In our model clusters, we use parameters and bond lengths typical 
for copper or nickel interacting with nitrogen and/or sulfur ligands. 
This is because copper and nickel have the largest ligand-field energies 
among the 3d transition metals, and typical ligands for these metals 
are nitrogen and sulfur.
%First ionization energy of S is 10.4 eV. This gives
%WAVE FUNCTION decay length of 0.605 AA, or
%0.303 AA for squared overlap.
We have examined four simple geometries;
the corresponding minimum-energy clusters are shown in Fig.~3.

a) A cluster with all four ligands $\sigma$-bonded and placed at 
equal distances of 2.0{\AA}. We take repulsion parameters appropriate 
for nitrogen. The result is a square-planar cluster ({\it cf.} Fig.~3a), 
consistent with the observed structures of many copper and nickel complexes.  
With this parameterization, the square structure is quite strongly favored 
over the tetrahedral one, by 0.31~eV or 7.0~kcal/mole. The square
structure remains stable under increases of the ligand-ligand
repulsion strengths. of up to a factor of over three.  We note, 
however, that the stability of the square structure is strongly 
sensitive to the choice of repulsive terms. For example, we find 
that when the van der Waals term from the OPLS-II force 
%field\cite{Jorgensen88} 
field~(Jorgensen et al., 1988) 
is used instead of the MM2 form, an increase of the repulsion strengths 
of only 20\% is enough to stabilize the tetrahedral structure. 
We expect that input from fully quantum-mechanical total-energy 
calculations will be necessary to precisely pin down the competition 
between the electronic energy and the repulsion terms in determining 
the structural energy differences. 

b) A cluster similar to that of a), but with one $\pi$-bonded ligand. 
This leads to a substantial deviation from planarity, 
as shown in Fig.~3b, where the $\pi$-bonded ligand is atom~4. 
As expected from the maximum of the $\pi$-$\sigma$ potential in Fig.~1, 
the angle between atom 4 and atoms 1--3 is greater than
$90^{\circ}$, about $123^{\circ}$. The angle between the
$\sigma$-bonded atoms 1, 2, and 3, is slightly greater than
the ideal value of $90^{\circ}$, because of the repulsive energy term.  

c) A cluster with with two $\sigma$-bonded nitrogen ligands 1 and 2, one
$\sigma$-bonded sulfur ligand 3, and a $\pi$-bonded sulfur ligand 4.
The distances are 2.04{\AA} for the nitrogen ligands, 
2.18{\AA} for the $\pi$-bonded sulfur ligand, and 
2.64{\AA} for the $\sigma$-bonded sulfur ligand. 
This geometry is motivated by the observed geometry of copper 
sites in ``blue-copper" proteins. In these proteins, 
the nitrogen ligands and a cysteine sulfur ligand are close 
to the copper, and it is believed that the cysteine sulfur 
ligand is primarily $\pi$-bonded. An additional $\sigma$-bonded
methionine sulfur ligand is farther from the copper. The values
of the ligand distances here are taken from cluster calculations
for blue-protein 
%%models\cite{Ryde96}. 
models~(Ryde et al., 1996). 
The van der Waals parameters are taken from the OPLS parameter set 
but we do not have a reliable procedure for determining the differences 
between the ligand-ion interactions of the nitrogen and the sulfur ligands.
Since the greater distance of the sulfur at 2.18{\AA} from the copper 
will be compensated to some extent by the greater size of sulfur 
relative to nitrogen, we simply assume that the three close ligands 
have the same coupling strength to the central atom. 
We have varied the ratio of the sulfur to nitrogen coupling strengths
by up to 30\% in both directions and found changes of only a few degrees 
in the bond angles of the cluster.  For the far ligand, we use the scaling 
procedure described above, which leads to a coupling that has 50\%  
of the strength of the close ligands. We have varied this coupling
from 0\% to 75\% of the close-ligand value, and again found bond-angle 
changes of only a few degrees. The lowest-energy structure for 
this cluster is indicated in Fig.~3c. The geometry is trigonal, 
with bond angles of $125^{\circ}$  between the cysteine sulfur 
and the nitrogens; the nitrogen-nitrogen bond angle is $91^{\circ}$. 
By comparison, the optimal values for protein models obtained by 
quantum-chemical 
%calculations\cite{Ryde96} 
calculations~(Ryde et al., 1996) 
are $125^{\circ}$ for the cysteine sulfur-nitrogen bond angle 
and  $103^{\circ}$ for the nitrogen-nitrogen bond angle. 
The results obtained by the present method are do of course not 
have quantitative accuracy, but the overall structure of the 
coordination shell is quite similar to that obtained in 
%%Ref. (\onlinecite{Ryde96}), 
(Ryde et al., 1996), 
which is shown in Fig.~3d. The stability of this structure is not 
predetermined by the bond lengths that we used as input. 
To demonstrate this, we have considered a cluster with the same 
bond lengths as the blue-protein fragment, with with all ligands 
$\sigma$-bonded.  In this case, the structure is planar, 
as shown in Fig.~3e. Thus the formation of the trigonal structure 
is directly related to the special character of the angular interactions 
associated with $\pi$-bonding.

\newpage
\section{Conclusion}

We have seen that it is possible to extract the functional form
of angular forces around transition metals in biomolecules by
using an approximate treatment of the ligand-field Hamiltonian. 
The approximations made here, involving primarily ionic bonding,  
are complementary to those used in 
%%Ref. (\onlinecite{Landis98}). 
(Landis et al., 1998). 
The resulting angular forms have simple trigonometric forms, 
which are very different for $\pi$-bonding as compared to $\sigma$-bonding. 
Ligand-field energies are represented well by these angular forces,
and much better than by force fields with assumed functional forms. 
Future work should aim to include the functional forms derived here
in widely used biomolecular simulation packages. 

\acknowledgments

This work was supported by the National Science Foundation under
Grant DMR-9971476.  We appreciate informative conversations with
Garland Marshall and Jay Ponder.

\newpage
%\bibliography{bpj}
%\begin{thebibliography}{10}
\centerline{\bf REFERENCES}

\begin{itemize}
\item{}
%\bibitem{Allured91}
Allured, V.~S., C.~M. Kelly, and C.~R. Landis. 1991.
Shapes Empirical Force Field:  New Treatment of Angular Potentials
and its Application to Square-Planar Transition-Metal Complexes.
{\it J. Am. Chem. Soc.} 113:1-12.
\item{}
%\bibitem{Carlsson98}
Carlsson, A.~E. 1998. Angular Forces Around Transition Metals
in Biomolecules. {\it Phys. Rev. Letters} 81:477-480.
\item{}
%\bibitem{Comba95}
Comba, P., T.~W. Hambley, and M. Str{\"o}hle. 1995.
The Directionality of d-Orbitals and Molecular-Mechanics
Calculations of Octhedral Transition-Metal Compounds.
{\it Helv. Chim. Acta} 78:2042-2047.
\item{}
%\bibitem{Cotton72}
Cotton, F.~A., and G. Wilkinson. 1972. Advanced Inorganic Chemistry.
Interscience, New York.
\item{}
%\bibitem{Jorgensen88}
Jorgensen, W.~L. and J. Tirado-Rives. 1988.
The OPLS Potential Functions for Proteins, Energy Minimizations
for Crystals of Cyclic Peptides and Crambin.
{\it J. Am. Chem. Soc.} 110:1657-1666.
%\bibitem{Kaim94}
\item{}
Kaim, W., and B. Schwederski. 1994. Bioinorganic Chemistry: Inorganic 
Elements in the Chemistry of Life. Wiley, New York.
\item{}
%\bibitem{Landis98}
Landis, C.~R., T. Cleveland, and T.~K. Firman. 1998.
Valence Bond Concepts Applied to the Molecular Mechanics
Description of Molecular Shapes.  3. Applications to Transition
Metal Alkyls and Hydrides.  {\it J. Am. Chem. Soc.} 120:2641-2649.
\item{}
%\bibitem{Ryde96}
Ryde, U., M.~H.~M. Olsson, K. Pierllot, and B.~O. Roos. 1996.
The Cupric Geometry of Blue Copper Proteins is not Strained.
{\it J. Mol. Biol.} 261:586-596.
\item{}
%\bibitem{Sayle95}
Sayle, D.~C., M.~A. Perrin, P. Nortier, and C.~R.~A. Catlow. 1995.
Simulation Study of Copper(I) and Copper(II) Species in ZSM-5
Zeolite.  {\it J. Chem. Soc. Chem. Commun.}  9:945-947.
\item{}
%\bibitem{Sayle97}
Sayle, D.~C., {\it et~al.}, 1997. 
Computer Modeling of the Active-Site Configurations within the NO
Decomposition Catalyst CU-ZSM-5. {J. Phys. Chem.} 101:3331-3337.
\item{}
%\bibitem{Sprague87}
Sprague, J.~T., J.~C. Tai, Y. Yuh, and N.~L. Allinger. 1987.
The MMP2 Calculational Method.  {\it J. Comput. Chem.} 8:581-603.
\item{}
%\bibitem{Timofeeva95}
Timofeeva, T.~V., J.~H. Lii, and N.~L. Allinger. 1995.
Molecular Mechanics Explanation of the Metallocene Bent Sandwich
Structure.  {\it J. Am. Chem Soc.} 117:7452-7459.
\item{}
%\bibitem{Wiesemann94}
Wiesemann, F., S. Teipel, B. Krebs, and U. Howeler. 1994.
Force-Field Calculations on the Structures of Transition-Metal
Complexes:  1. Application to Copper(II) Complexes in
Square-Planar Coordination.  {\it Inorg. Chem.} 33:1891-1898.
\item{}
%\bibitem{Wigner59}
Wigner, E.~P. 1959. Group Theory and its Application to the Quantum 
Mechanics of Atomic Spectra. Academic Press, New York.
\end{itemize}
%\end{thebibliography}
%\end{thebibliography}
\newpage
\begin{figure}
\caption{ (a) Interaction between ligands of central transition metal
ion. The dimensionless quantity $u$ ({\it cf.} Eqs.~(19)--(21))
is either $u_{\sigma,\sigma}$ (solid line), 
$u_{\pi,\sigma}$ (dashed line), or $u_{\pi,\pi}$ (dotted line).
The quantity $-u$ is plotted so that negative values will 
correspond to lower energies.
(b) Empirical force field from (Comba et al., 1995).}
\end{figure}

\begin{figure}
\caption{Energies obtained by semiclassical force field vs.
exact energies for small model clusters. Energies given in 
units of $h_{\rm max}$, maximal coupling strength between ligands and
transition metal. (a)~One $\pi$-bonded ligand and three
$\sigma$-bonded ligands. (b)~Four $\pi$-bonded ligands.}
\end{figure}

\begin{figure}
\caption{Lowest-energy structures obtained in relaxations of
five-atom cluster using angular forces. (a)~Four equivalent
$\sigma$-bonded neighbors.  (b)~Three $\sigma$-bonded neighbors
and one $\pi$-bonded neighbor~4, all at the same distance.
(c)~Two $\sigma$-bonded nitrogen neighbors~1 and 2, a near
$\pi$-bonded sulfur neighbor~3, and a far $\sigma$-bonded
neighbor~4.  (d)~Structure for blue protein model obtained by
quantum-chemical calculations (Ryde et al., 1996). (e)~Three near
$\sigma$-bonded neighbors and a far $\sigma$-bonded neighbor.}
\end{figure}

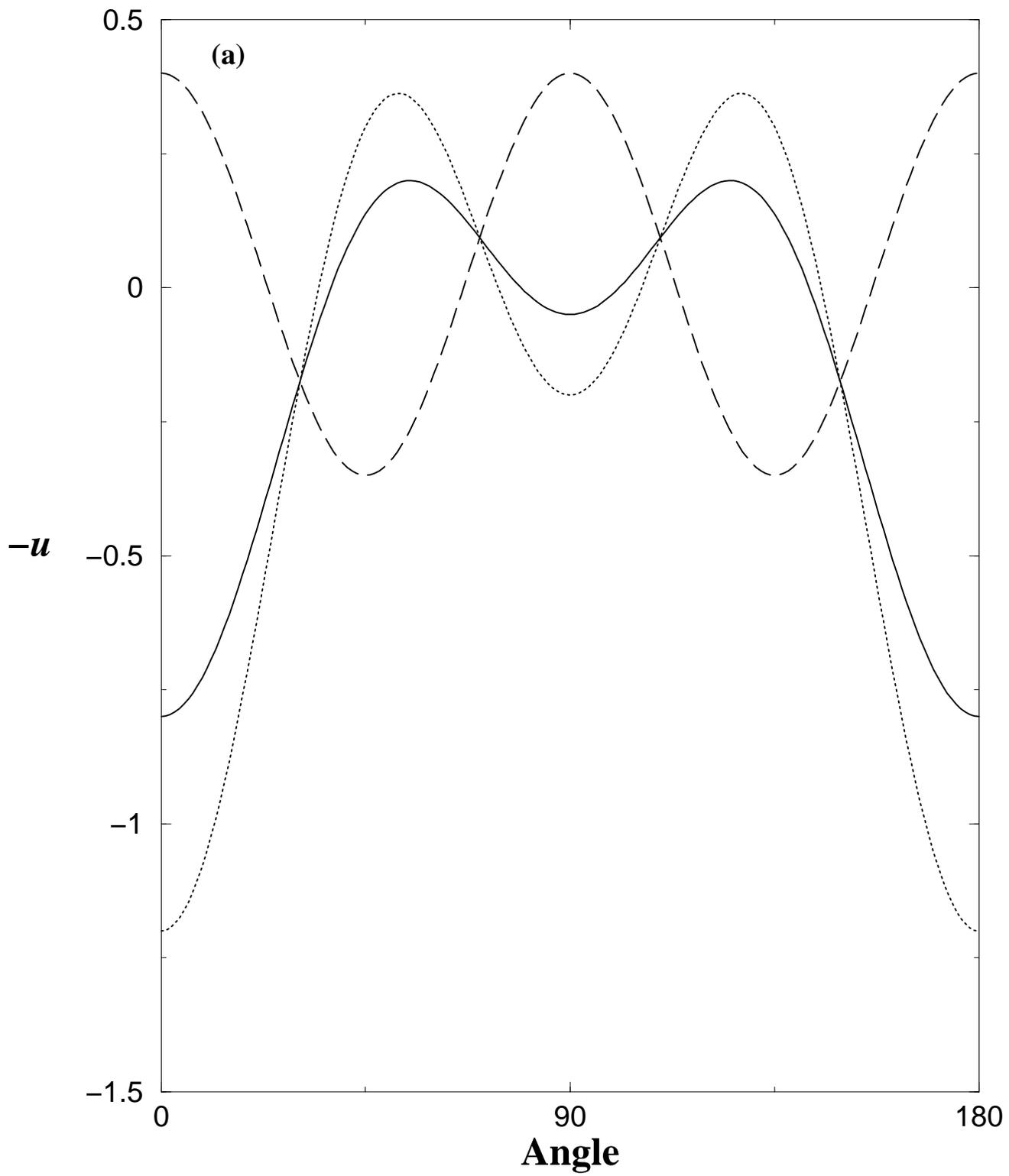
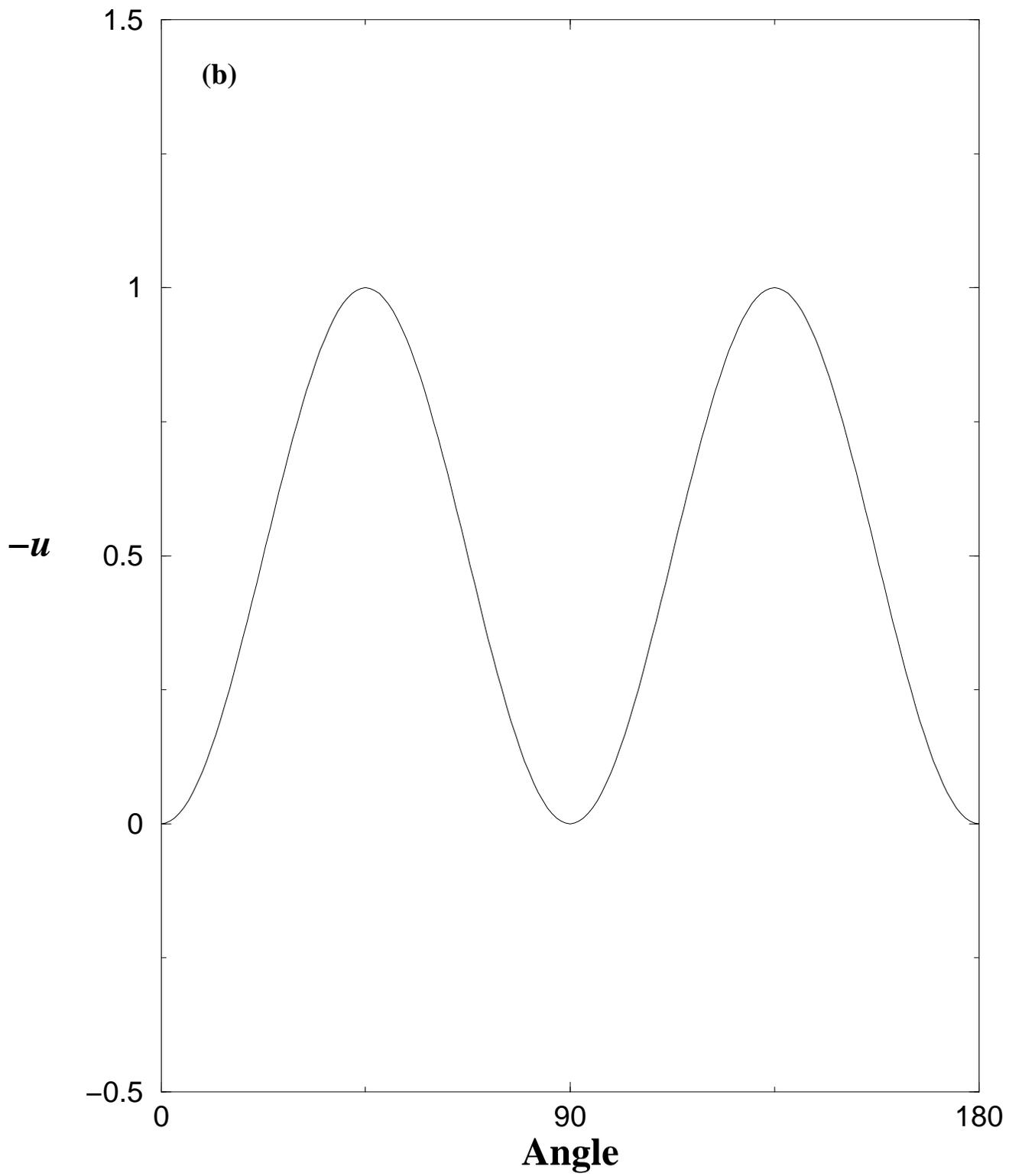
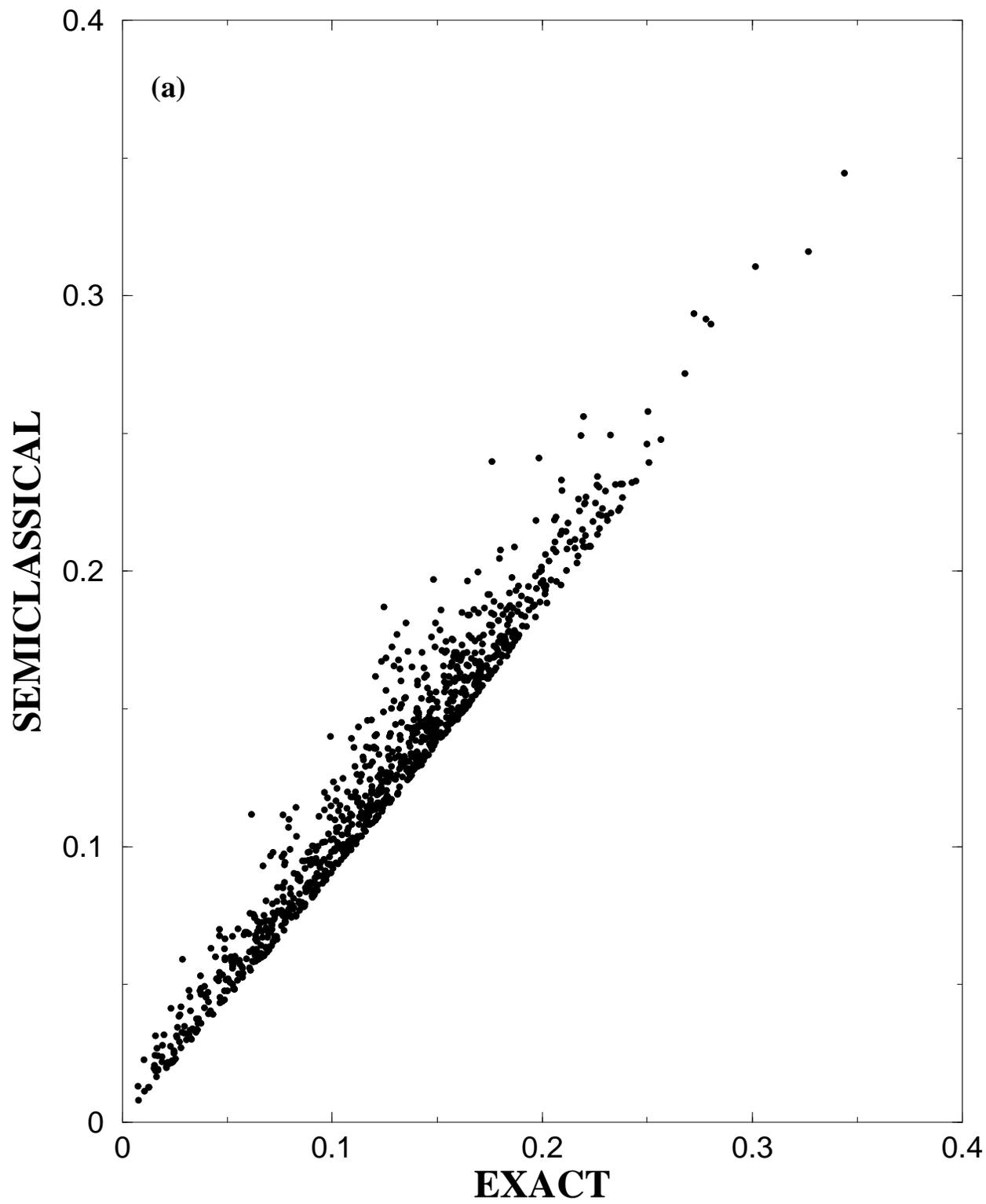
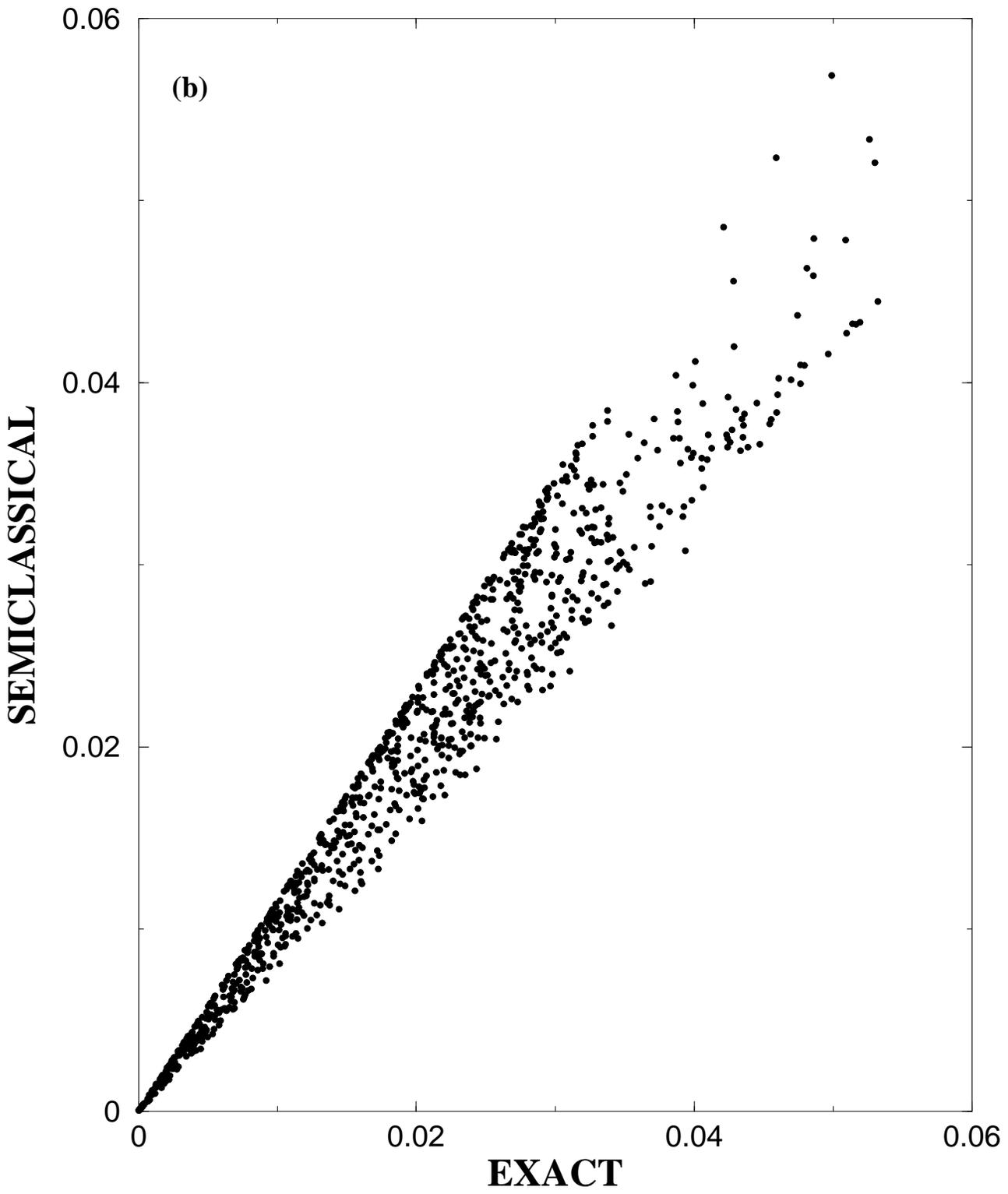
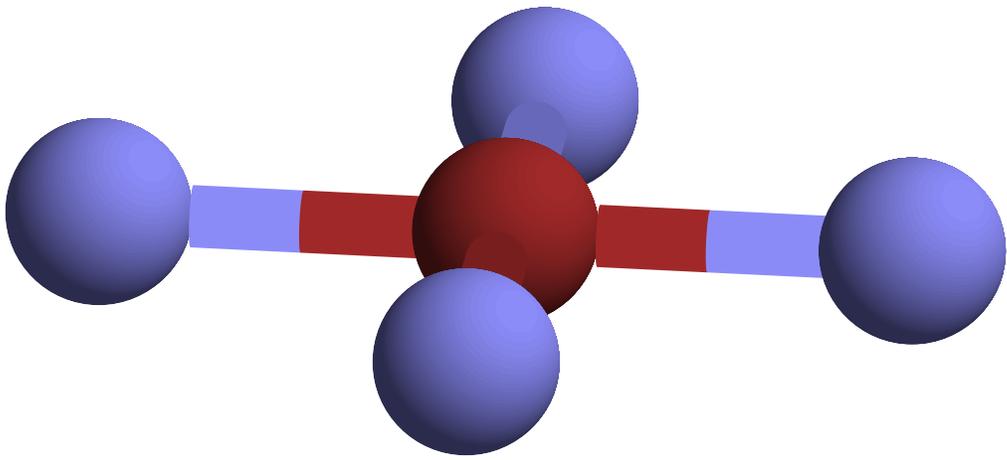
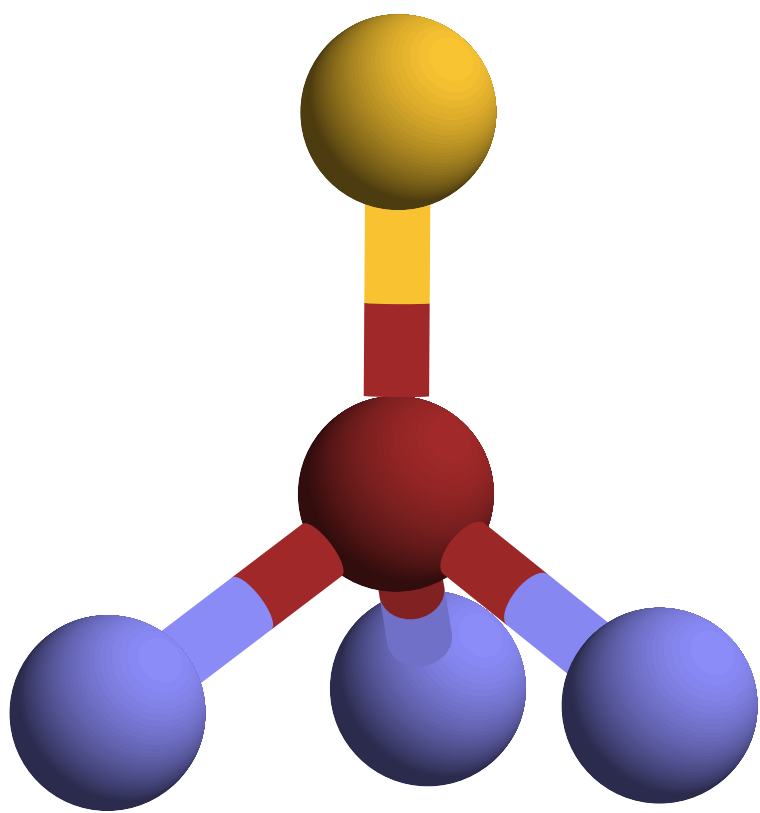
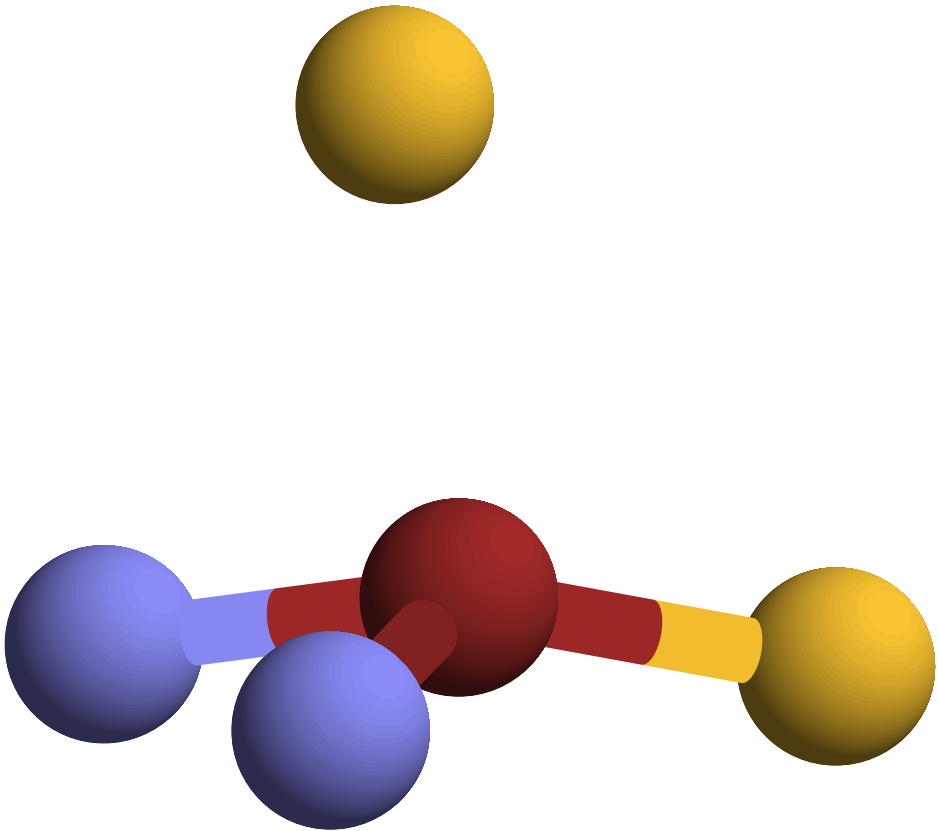
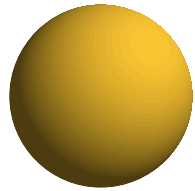
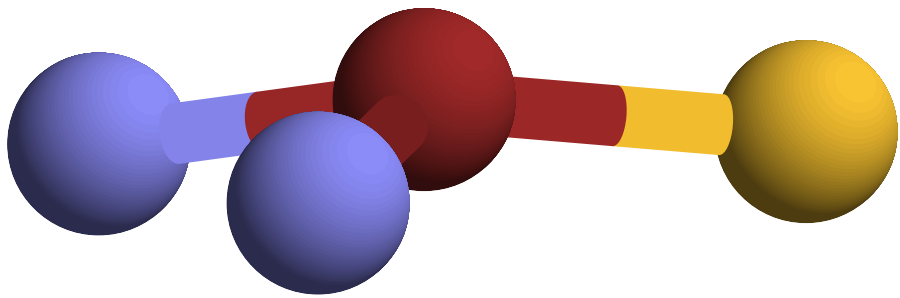
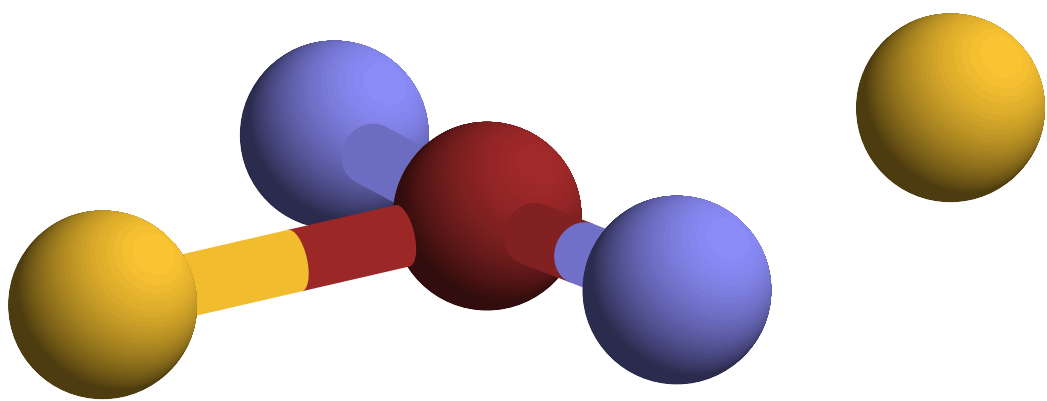
\end{document}